# Transverse spinning of unpolarized light

J. S. Eismann[1,2,*], L. H. Nicholls[3,*], D. J. Roth[3,*], M. A. Alonso[4,5], P. Banzer[1,2,#], F. J. Rodríguez-Fortuño[3], A. V. Zayats[3,#], F. Nori[6,7], and K. Y. Bliokh[6,#]

[1]*Max Planck Institute for the Science of Light, Staudtstrasse 2, D-91058 Erlangen, Germany*
[2]*Institute of Optics, Information, and Photonics, University Erlangen-Nuremberg, Staudtstrasse 7/B2, D-91058 Erlangen, Germany*
[3]*Department of Physics and London Centre for Nanotechnology, King's College London, Strand, London WC2R 2LS, UK*
[4]*CNRS, Centrale Marseille, Institut Fresnel, Aix Marseille Univ, UMR 7249, 13397 Marseille CEDEX 20, France*
[5]*The Institute of Optics, University of Rochester, Rochester, NY 14627, USA*
[6]*Theoretical Quantum Physics Laboratory, RIKEN Cluster for Pioneering Research, Wako-shi, Saitama 351-0198, Japan*
[7]*Physics Department, University of Michigan, Ann Arbor, Michigan 48109-1040, USA*
[*]*These authors contributed equally to this work*
[#]*Corresponding authors*

It is well known that spin angular momentum of light, and therefore that of photons, is directly related to their circular polarization. Naturally, for totally unpolarized light, polarization is undefined and the spin vanishes. However, for nonparaxial light, the recently discovered *transverse spin* component, orthogonal to the main propagation direction, is largely independent of the polarization state of the wave. Here we demonstrate, both theoretically and experimentally, that this transverse spin survives even in nonparaxial fields (e.g., tightly focused or evanescent) generated from a *totally unpolarized* light source. This counterintuitive phenomenon is closely related to the fundamental difference between the degrees of polarization for 2D paraxial and 3D nonparaxial fields. Our results open an avenue for studies of spin-related phenomena and optical manipulation using unpolarized light.

## 1. Introduction

Classical polarization optics usually regards paraxial light and its 2D polarization states [1]. Similarly, the spin of photons in quantum electrodynamics textbooks is also described by 2D circular polarizations of plane electromagnetic waves [2]. However, modern nano-optics is based on the use of structured nonparaxial fields, where all three spatial components of the field vector generically play a role [3]. This required extending the existing polarization theory to the 3D case [4–8]. This extension is by no means trivial: the four Stokes parameters describing generic 2D polarization are now substituted by *nine* polarization parameters characterizing generic 3D polarization.

Simultaneously, the notion of spin has to be augmented to 3D structured fields [9–13], where the local spin density is well-defined for monochromatic waves and can be associated with the radiation torque on small dipole particles [13]. This resulted in the discovery of the unusual *transverse spin* in inhomogeneous fields with several remarkable properties [14–28] (for review, see [13,29–31]). This spin, orthogonal to the main propagation direction and wavevectors, is a very robust phenomenon that has found applications for highly efficient spin-direction coupling using evanescent waves, largely independently of the details of the system [18–21,23–25,29–31]. Moreover, it was recently found that the transverse spin is equally present in inhomogeneous sound waves [32–34], which are traditionally considered as scalar (i.e., spinless), quantum electron waves [25], and even gravitational waves [35].

In this work, we demonstrate, both theoretically and experimentally, that the transverse spin is essentially a polarization-independent phenomenon, which survives even in fields generated by *totally unpolarized* sources, Fig. 1. This is in sharp contrast to the usual longitudinal spin, which is directly related to the 2D polarization and vanishes in unpolarized fields. We show that this phenomenon is intimately related to the difference between the 2D and 3D polarization descriptions. Namely, the totally depolarized 2D field is at the same time *half-polarized* in the 3D sense [5]. Indeed, 2D depolarization implies a *single* random phase between the two orthogonal field components (with equal amplitudes), while complete 3D depolarization requires *two* random phases between the three mutually-orthogonal field components. Therefore, any regular optical transformation producing a nonparaxial 3D field from an unpolarized far-field source will have partial 3D polarization, with the degree of polarization not less than 1/2. In particular, the local increase of the degree of polarization up to almost 1 was demonstrated for the tight focusing of an unpolarized paraxial beam [36,37]. Below we show that the transverse spin appears in any paraxial-to-nonparaxial transformation (see Fig. 1), even without a change in the degree of polarization; the minimal value of 1/2 provides sufficient room for this. The origin of this phenomenon lies in intrinsic *spin-orbit interaction* of light [30], where any transformation in the wavevector direction produces spin-related phenomena, even for unpolarized light.

Since spin is a fundamental dynamical property of light, which is very important in both quantum and classical, theoretical and applied optics (e.g., for optical manipulation of micro- and nano-particles), our findings provide a novel opportunity to use polarization-independent spin from unpolarized sources.

## 2. Theoretical background

Nonparaxial optical fields are usually generated from far-field sources of paraxial light via some optical transformations (see Fig. 1): focusing, diffraction, scattering, etc. In this work, we consider two of the most common examples of nonparaxial fields: (i) tightly focused Gaussian-like beams and (ii) evanescent waves. These are generated via high-NA focusing and total internal reflection of the incident paraxial light, respectively.

The incident paraxial light can be approximated by a plane wave, so its 2D polarization state can be described by the 2×2 polarization (density) matrix or, equivalently, by 4 real Stokes parameters $\vec{s} = (s_0, s_1, s_2, s_3)$ [1]. Here, the normalized parameter $s_3$ corresponds to the normalized spin angular momentum density of the wave ($z$-directed along the wave propagation): $S_z / I = s_3 / s_0 \in [-1, 1]$ [13], where $I = W / \omega$ is the wave intensity expressed via the energy density $W$ and frequency $\omega$. The degree of paraxial 2D polarization is defined as $P^{2D} = \sqrt{\sum_{i=1}^{3} s_i^2} / s_0 \in [0, 1]$. For totally unpolarized light, $\vec{s} \propto (1, 0, 0, 0)$, $P^{2D} = 0$, and the spin vanishes: $S_z = 0$ (see Fig. 1).

For the generated nonparaxial field, all three components are significant, and its polarization state at a point is described by a 3×3 Hermitian polarization (density) matrix, or equivalently by 9 real parameters $\vec{\Lambda} = (\Lambda_0, \Lambda_1, ..., \Lambda_8)$ [4–8] [see Supplementary Information (SI)]. In such fields, the polarization ellipsoid can have an arbitrary orientation, and the spin angular momentum density (orthogonal to it) involves all three components [10,13]. Its normalized value can be expressed via the properly normalized parameters $\Lambda_2$, $\Lambda_5$, and $\Lambda_7$ (see SI):

$$\frac{\mathbf{S}}{I} \equiv \frac{1}{I}\left(S_x, S_y, S_z\right) = \frac{2}{3\Lambda_0}\left(-\Lambda_7, \Lambda_5, -\Lambda_2\right). \quad (1)$$

The most common definition of the degree of nonparaxial 3D polarization is $P^{3D} = \sqrt{\sum_{i=1}^{8} \Lambda_i^2}/\sqrt{3}\Lambda_0 \in [0,1]$ [4–8]. For totally unpolarized 3D light, one should expect $\vec{\Lambda} \propto (1,0,0,...,0)$, $P^{3D} = 0$, and the corresponding vanishing spin: $\mathbf{S} = \mathbf{0}$.

One remarkable feature of the above definitions of the degree of polarization is that totally unpolarized paraxial light, $P^{2D} = 0$, is *partially* polarized in the 3D sense: $P^{3D} = 1/2$ [5] (see SI). This is because total 3D depolarization requires total mutual decoherence of all of the three field components with equal amplitudes, while in paraxial light the longitudinal $z$ component vanishes, and in fact remains "coherent" with the mutually decoherent transverse $(x, y)$ components. As a result, $\Lambda_8 = \sqrt{3}/2\Lambda_0 \neq 0$ even for a totally unpolarized paraxial field. This "discrepancy" between the 2D and 3D polarization degrees naturally manifests itself as a *nonzero transverse spin* in a nonparaxial field generated from an unpolarized paraxial source, Fig. 1.

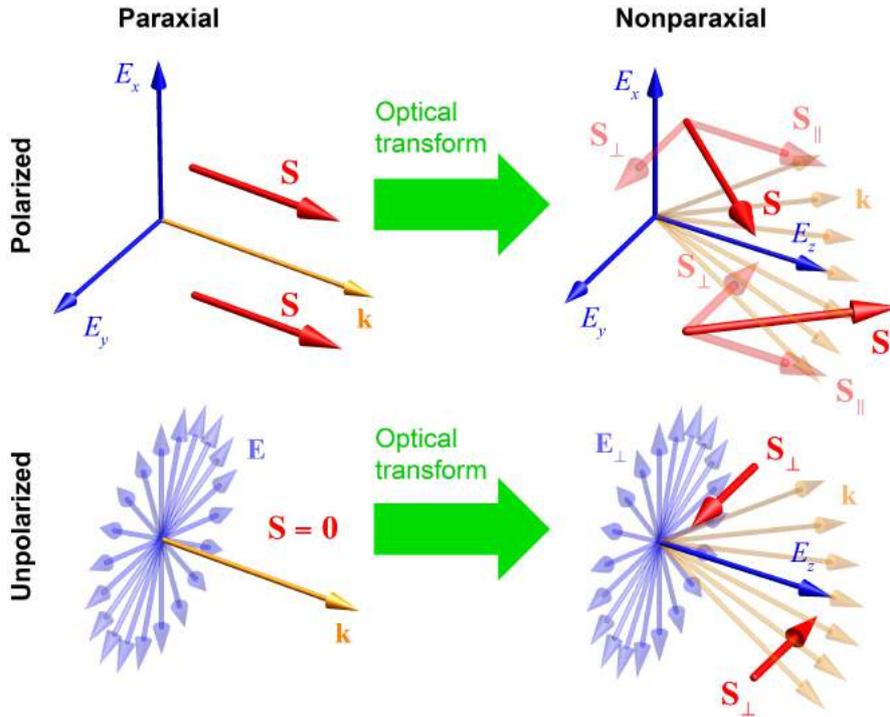

**Figure 1**. Schematic illustration of the longitudinal and transverse spin for the paraxial and nonparaxial regimes for both polarized and unpolarized (in the 2D sense) fields. Transverse spin $\mathbf{S}_\perp$ appears in nonparaxial fields, while the depolarization of the paraxial source eliminates only the longitudinal spin $\mathbf{S}_\parallel$.

We first consider the case of a tightly focused polarized field. Both the incident paraxial and focused nonparaxial fields can be modeled by the post-paraxial description of a Gaussian beam [13] with the infinite and finite Rayleigh range $z_R$, respectively. Using the natural cylindrical coordinates $(r, \varphi, z)$, the normalized spin density in a polarized Gaussian beam can be written as (see SI):

$$\frac{\mathbf{S}}{I} \simeq \frac{1}{1+\tilde{r}^2/2}\left[\frac{\mathbf{S}_0}{I_0} + \tilde{r}\overline{\boldsymbol{\varphi}}\right] \equiv \frac{\mathbf{S}_\parallel}{I} + \frac{\mathbf{S}_\perp}{I}. \tag{2}$$

Here $\mathbf{S}_0/I_0 = (s_3/s_0)\overline{\mathbf{z}}$ is the spin density in the plane-wave limit, $I \propto (1+\tilde{r}^2/2)e^{-kr^2/z_R}$ is the intensity distribution, $\tilde{r} = r/z_R$, and the overbars indicate the unit vectors of the corresponding axes. Equation (2) exhibits the usual polarization-dependent longitudinal spin, as well as the transverse spin component [13,22,28,29] which is totally independent of the polarization (Stokes parameters) of the incident plane wave.

The totally unpolarized Gaussian beam can be considered as an *incoherent* superposition of two Gaussian beams with mutually orthogonal polarization states (e.g., with $\vec{s} \propto (1,1,0,0)$ and $\vec{s} \propto (1,-1,0,0)$). The corresponding 3×3 polarization matrix and parameters $\vec{\Lambda}$ for such unpolarized Gaussian field become (see SI): $\Lambda_1 = \Lambda_2 = \Lambda_3 = \Lambda_4 = \Lambda_6 = 0$,

$$\frac{\Lambda_8}{\Lambda_0} = \frac{\sqrt{3}}{2}\frac{1-\tilde{r}^2}{1+\tilde{r}^2/2}, \quad \frac{\Lambda_5}{\Lambda_0} = \frac{3}{2}\frac{\tilde{x}}{1+\tilde{r}^2/2}, \quad \frac{\Lambda_7}{\Lambda_0} = \frac{3}{2}\frac{\tilde{y}}{1+\tilde{r}^2/2}, \tag{3}$$

where $\tilde{x} = x/z_R$ and $\tilde{y} = y/z_R$. In the paraxial limit $z_R \to \infty$, only the $\Lambda_8/\Lambda_0$ ratio survives, providing the 3D degree of polarization $P^{3D} = 1/2$ [5]. In the nonparaxial case, the nonzero parameters $\Lambda_5$ and $\Lambda_7$ appear. These parameters exactly describe the transverse part of spin (2) in agreement with Eq. (1): $\frac{\mathbf{S}_\perp}{I} = \frac{2}{3\Lambda_0}(-\Lambda_7, \Lambda_5, -\Lambda_2)$, while the longitudinal spin naturally vanishes: $\mathbf{S}_\parallel = \mathbf{0}$ (see Fig. 1).

Second, we consider an evanescent wave, which can be generated via total internal reflection of a paraxial incident field (plane wave). Such *z*-propagating and *x*-decaying wave is characterized by the propagation constant $k_z > k \equiv \omega/c$ and the decay constant $\kappa = \sqrt{k_z^2 - k^2}$. Assuming, for simplicity, that the transmission coefficients of the total internal reflection are polarization-independent, the generation of the evanescent field can be regarded as a transition from the plane-wave limit $\kappa = 0$, $k_z = k$, to the given finite $\kappa > 0$. The normalized spin density of the polarized evanescent wave is [13,16]:

$$\frac{\mathbf{S}}{I} = \frac{k}{k_z}\frac{\mathbf{S}_0}{I_0} + \frac{\kappa}{k_z}\overline{\mathbf{y}} \equiv \frac{\mathbf{S}_\parallel}{I} + \frac{\mathbf{S}_\perp}{I}. \tag{4}$$

Here, as before, $\mathbf{S}_0/I_0 = (s_3/s_0)\overline{\mathbf{z}}$ is the spin density in the plane-wave limit, and the intensity distribution is $I \propto e^{-2\kappa x}$. As for the focused field, the spin (4) consists of the longitudinal polarization-dependent component and transverse (*y*-directed) polarization-independent term [13,16,25,29,30].

The totally unpolarized evanescent field is obtained as an incoherent superposition of evanescent waves with orthogonal polarization states. The corresponding parameters $\vec{\Lambda}$ for such unpolarized evanescent field are (see SI): $\Lambda_1 = \Lambda_2 = \Lambda_4 = \Lambda_6 = \Lambda_7 = 0$,

$$\frac{\Lambda_8}{\Lambda_0} = \frac{\sqrt{3}}{2}\frac{k^2 - \kappa^2/2}{k_z^2}, \quad \frac{\Lambda_3}{\Lambda_0} = \frac{3}{4}\frac{\kappa^2}{k_z^2}, \quad \frac{\Lambda_5}{\Lambda_0} = \frac{3}{2}\frac{\kappa}{k_z}. \tag{5}$$

In the plane-wave limit $\kappa = 0$, only the ratio $\Lambda_8 / \Lambda_0$ survives, yielding $P^{3D} = 1/2$. In the evanescent-wave case, both $\Lambda_3$ and $\Lambda_5$ are different from zero, the latter corresponding precisely to the transverse part of spin (4) in agreement with Eq. (1): $\dfrac{\mathbf{S}_\perp}{I} = \dfrac{2}{3\Lambda_0}\left(-\Lambda_7, \Lambda_5, -\Lambda_2\right)$, whereas the longitudinal spin vanishes: $\mathbf{S}_\parallel = \mathbf{0}$ (see Fig. 1).

Importantly, considering $r/z_R$ and $\kappa/k$ as a small parameter $\varepsilon$ in the above two problems, the 3D degree of polarization of the unpolarized focused and evanescent fields has the form $P^{3D} = \dfrac{1}{2} + O(\varepsilon^2)$ (see SI), while the transverse spin has the order of $\varepsilon$. This means that, to first order, focusing or total-reflection processes (with polarization-independent transmission amplitudes) do not change the 3D degree of polarization of the incident unpolarized light [36,37], while the spin changes from zero in the incident wave to the nonzero transverse spin in the nonparaxial field. This appearance of spin without polarization originates from the intrinsic *spin-orbit interaction* of light [30]. The plane-wave transversality condition $\mathbf{k} \cdot \mathbf{E} = \mathbf{k} \cdot \mathbf{H} = 0$ imposes constraints on the relations between longitudinal and transverse field components, which therefore have some intrinsic mutual coherence even for fields generated from unpolarized sources. Transformations from paraxial to nonparaxial field can be approximated by $\mathbf{k}$-vector transformations (re-directions), which do not affect the polarization degree but inevitably generate the transverse spin, as schematized in Fig. 1.

Another important point is that in our calculations we considered both *electric* and *magnetic* field contributions to all quadratic quantities (see SI): spin, intensity, polarization parameters $\vec{\Lambda}$, etc. For polarized fields, the electric and magnetic contributions are not equal to each other, and additional terms generally appear when considering only the electric or the magnetic fields [13,16,22,28]. In contrast, for unpolarized fields, these contributions are always equal to each other. One can say that unpolarized light and its transverse spin have *dual-symmetric* nature [12,38], similarly to circularly-polarized fields with well-defined helicity [11].

In what follows, we present experimental measurements of the nonzero transverse spin from Eqs. (2) and (4) in tightly focused and evanescent fields generated from unpolarized sources. The two experiments use different types of unpolarized sources and measure both the electric and magnetic contributions to the spin.

## 3. Focused-beam experiment

In order to measure the transverse spin of an unpolarized tightly focused beam, we first prepared a suitable input field. We sent a Gaussian beam (wavelength $\lambda = 2\pi/k = 620\,\text{nm}$, linewidth $\Delta\lambda_{\text{FWHM}} \simeq 5\,\text{nm}$) through a linear polarizer and two liquid-crystal variable retarders (LCs) oriented at 45° and 90° with respect to the axis of the linear polarizer, respectively. The experimental setup is schematically shown in Fig. 2a [28,39]. With this arrangement, the polarization state of the generated beam can span the whole Poincaré sphere ($\sum_{i=1}^{3} s_i^2 = s_0^2$) with the position on the sphere depending on the settings of the LCs. These LCs were controlled via a voltage applied to the corresponding devices to induce a voltage-dependent birefringence. For the applied voltage, we used two random numbers in a range spanning multiple wavelengths of retardance, updated 10 times per second. This produced a beam that is fully and homogeneously polarized over its cross-section for a fixed instance in time. However, the beam appears totally unpolarized ($P^{2D} = \sum_{i=1}^{3} s_i^2 = 0$) when averaged over a certain time frame.

For tight focusing and subsequent collimation of the light beam, we used two confocally aligned microscope objectives (MOs) with numerical apertures $NA_1 = 0.9$ and $NA_2 = 1.3$, respectively (see Fig. 2a). Following a scheme developed recently [28] for the reconstruction of the electric and magnetic parts of the transverse spin, we used a spherical silicon nanoparticle of diameter $d = 168$ nm as a local probe in the focal volume. The NA of the collection $MO_2$ was considerably larger than 1 in order to access the angular range above the critical angle, which is required for the applied reconstruction technique. Then, we performed a polarization analysis in the back focal plane (BFP) of $MO_2$ imaged onto a camera, which allowed us to access the far field of the scattered light. This polarization analysis involved a LC, a linear polarizer and an imaging lens (see Fig. 2a). At this stage of the setup, a single LC was sufficient because for the reconstruction of the transverse spin we only need to distinguish between the *x*- and *y*-polarizations. According to the method in Ref. [28], intensities of the *x*- and *y*-components of the scattered field, dependent on the transverse wavevectors, $I^{sc}_{x,y}(\mathbf{k}_\perp)$, allowed unambiguous reconstruction of both the electric and magnetic field contributions to the transverse spin density, $\mathbf{S}^{(e)}_\perp$ and $\mathbf{S}^{(m)}_\perp$, in the focused field at the location of the particle.

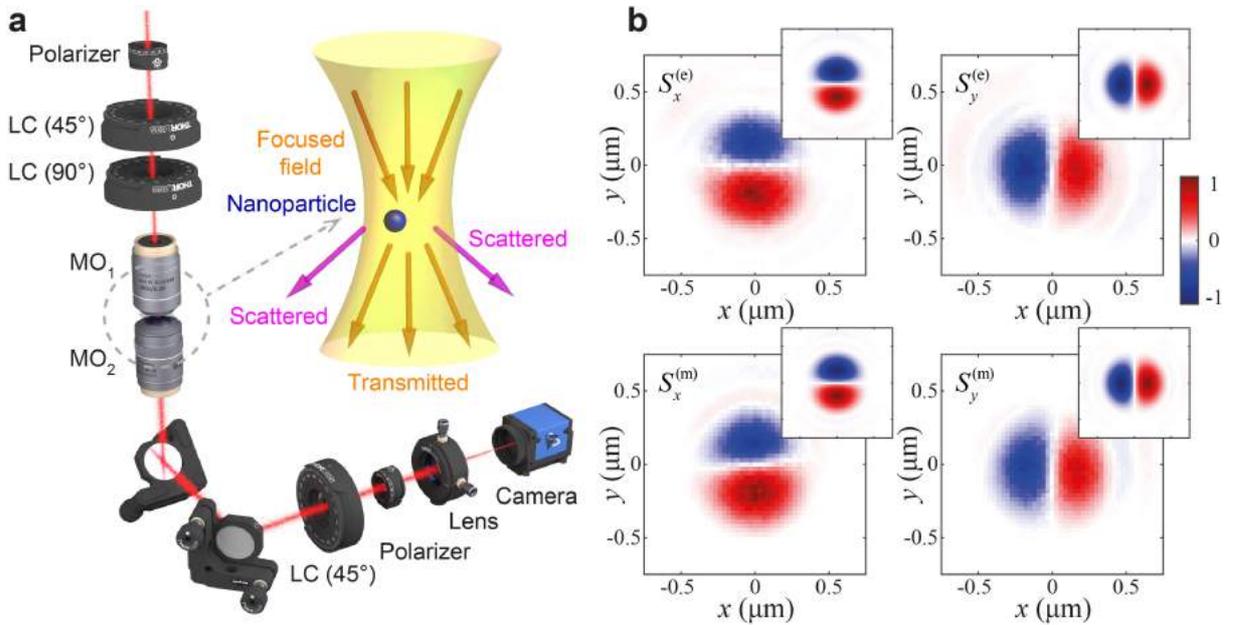

**Figure 2**. **a**, Experimental setup for the reconstruction of the transverse spin in a tightly focused unpolarized field. A linear polarizer and two liquid crystal variables retarders (LCs) are used to prepare a beam with randomly varied polarization. Subsequently, two confocally aligned microscope objectives (MOs) focus and collimate the beam. A spherical silicon nanoparticle is placed on a coverslip in the focal plane. It produces scattered light with wavevectors outside of the aperture of the transmitted beam, which carries information about the local transverse spin density in the beam [22,28]. Polarization-resolved back focal plane images using the scattered light are recorded by using another LC, a linear polarizer and a lens. **b**, Experimental results of the reconstructed electric and magnetic transverse spin, $\mathbf{S}^{(e)}_\perp$ and $\mathbf{S}^{(m)}_\perp$ (normalized to the maximum absolute value), which equal each other in the unpolarized field (see SI). The results of numerical calculations are shown as insets.

In order to provide a full map of the transverse electric and magnetic spin densities, $\mathbf{S}^{(e)}_\perp$ and $\mathbf{S}^{(m)}_\perp$, shown in Fig. 2b, we raster scan the nanoparticle across the focal plane (over the square area of $1.5 \times 1.5\,\mu m^2$ with a step size of $30\,nm$) and record the polarization-resolved BFP images for each particle position. For each position and polarization, the data is averaged over a time frame of $40\,s$. The distributions of the transverse spin obtained experimentally are in good agreement with simple theoretical expression (2) with the fitted Rayleigh range $z_R \simeq 527\,nm$. We also performed more accurate numerical calculations of the transverse spin densities using vectorial diffraction theory [40] (which takes into account the finite aperture of the focused beam) and plotted these as insets in Fig. 2b. In doing so, we adjusted all parameters of the focusing system and the incoming beam to the experimental case. One can see that the experimental results are in excellent agreement with the numerical data.

Importantly, the electric and magnetic spin densities in Fig. 2b exhibit very similar spatial distributions, in agreement with the dual-symmetric nature of the transverse spin for unpolarized light: $\mathbf{S}_\perp^{(e)} = \mathbf{S}_\perp^{(m)} = \mathbf{S}_\perp / 2$ (see SI). The same feature is present in nonparaxial fields with well-defined helicity [11], such as fields obtained by focusing circularly polarized input light [41]. However, in our case of an unpolarized source, the helicity and longitudinal spin vanish.

## 4. Evanescent-wave experiment

In order to measure the transverse spin of an unpolarized evanescent wave, the total internal reflection of light coming from an unpolarized tungsten lamp was employed. To generate the evanescent wave, a BK7 glass prism (Thorlabs, refractive index $n = 1.51$ at the wavelength $\lambda = 600\,nm$) was illuminated by an unpolarized tungsten lamp of wavelength 500–800 nm. The angle of incidence was measured to be 49°, which changes to 47° upon refraction entering the right-angle prism. This is above the critical angle of 41°, which is required for total internal reflection generating an evanescent wave above the glass. Akin to the focused-beam experiment, a small nanoparticle acting as a probe for the local field polarization – in this case gold nanoparticle (diameter $d = 150\,nm$, Sigma Aldrich) – was placed in the evanescent field above the prism and the far-field scattered radiation was analyzed (see Fig. 3a).

The scattered signal from the gold nanoparticle was collected by a 100× microscope objective with a numerical aperture $NA = 0.9$, allowing us to analyze the scattered light within a very large solid angle. The BFP of the detection objective (Fourier plane) was then imaged onto an imaging spectrometer using a set of relay lenses. The scattered signal was analyzed using a linear polarizer and a quarter wave plate in order to reconstruct the full Stokes parameters of the light scattered from the particle in all directions in the upper half-space (see SI). Figure 3b shows the results of these measurements, i.e., angular dependences of the normalized Stokes parameters $s_{1,2,3}/s_0$, as well as the 2D degree of polarization, $P^{2D}$, for the far-field scattering from the nanoparticle.

Note that the gold nanoparticle in this experiment behaves as an *electric* dipole, i.e., it is sensitive to the electric rather than magnetic field properties. However, we have already shown that the magnetic field shares the same features in unpolarized light, so we omit the superscript "(e)".

The degree of polarization $P^{2D}$ and third Stokes parameter $s_3/s_0$ in the scattered radiation show that the scattered light becomes partially polarized and acquires opposite-sign spins in the $\pm y$ directions. This is in perfect agreement with the $y$-directed transverse spin in Eq. (4) and the well-established fact that this transverse spin in an evanescent field is converted to the usual far-field spin (i.e., the third Stokes parameter) upon transverse scattering by a dipole particle [19–21,23–25,29–31]. The insets in Fig. 3b show the analytically calculated Stokes parameters of the scattered light for an unpolarized $\lambda = 600\,nm$ source. (The patterns depend very weakly on wavelength so that they are almost constant within the whole 500–800 nm range.) The analytical calculation was performed by matching the experimental parameters (angle of incidence, type of glass, particle diameter and material), including the total internal reflection of the incident beam, the particle modeled as a point dipole, and the subsequent scattering of the particle (taking into account the effects of the surface reflections; see SI). One can see a very good agreement between the theory and the experiment.

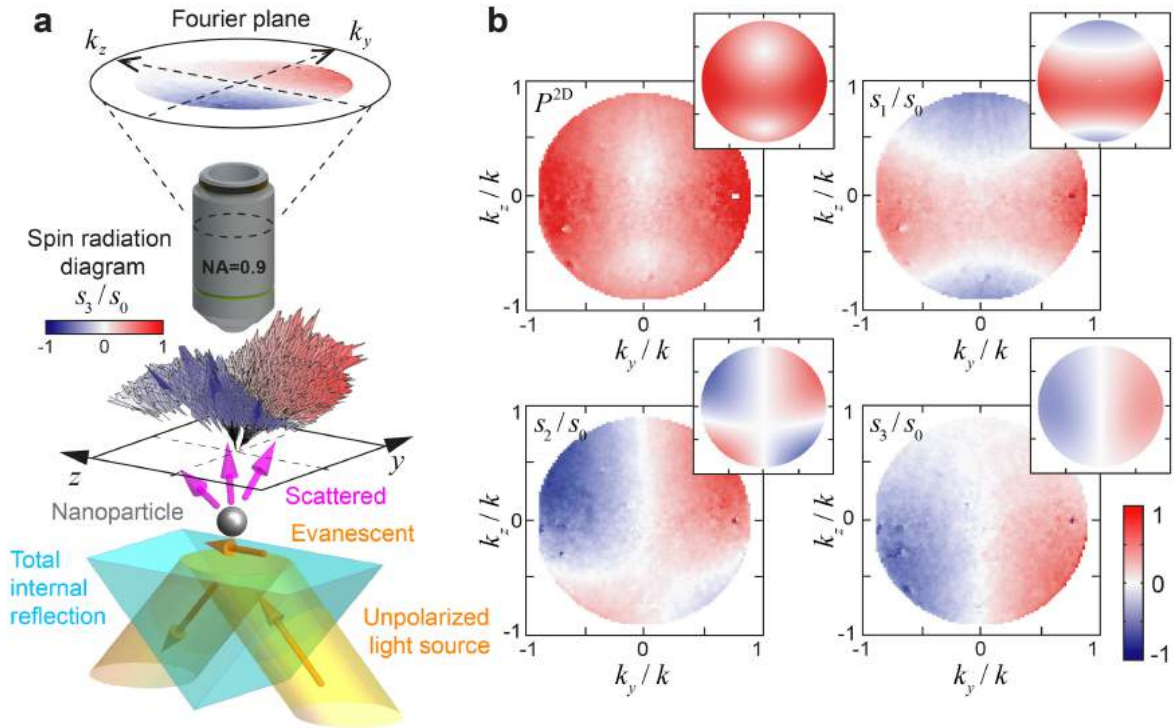

**Figure 3**. **a**, Experimental setup used to detect the non-zero transverse spin in an evanescent wave from an unpolarized source. Light from an unpolarised source undergoes total internal reflection, generating an evanescent wave, which is then scattered by a nanoparticle. The scattering from this nanoparticle is collected via a microscope objective. The radiation diagram above the nanoparticle represents the measured $P^{2D}$ (i.e. the degree of polarization in different directions), whereas the color represents the spin of the far-field radiation given by $s_3/s_0$. **b**, Experimentally retrieved and analytically calculated (inset) maps of $P^{2D}$ and normalized Stokes parameters $s_{1,2,3}/s_0$ of the scattered light in every direction of the upper half-space.

## 5. Conclusions

We have shown that pure redirection of wavevectors can generate nonzero spin angular momentum in initially completely unpolarized paraxial light. This surprising result establishes an important link between two areas of research: (i) 3D polarization in nonparaxial fields [4–8,36,37] and (ii) transverse spin [13–35]. The direct relation between the redirection of wavevectors and the appearance of spin points to the fundamental spin-orbit interaction origin of this phenomenon [30]. We have provided theoretical calculations and two sets of experimental measurements for the transverse spin generated upon tight focusing and total internal reflection (i.e., generation of an evanescent wave) of an unpolarized paraxial light. All these results use well-established methods for spin calculations and measurements, and are in perfect mutrual agreement.

Thus, our work has revealed one more exceptional feature of transverse spin. Together with other properties found previously, we can conclude that transverse spin is not just "one of the components of spin angular momentum density", but rather a separate physical entity whose main features are completely different from those of the usual polarization-controlled longitudinal spin of paraxial light or photons. As such, the transverse spin can offer novel phenomena and applications in angular-momentum and polarization optics. The remarkable "spin-momentum locking" associated with the transverse spin has already found promising applications for highly efficient spin-direction couplers [18–21,23–25,29–32]. The present study opens an avenue for the use of spin from unpolarized and incoherent sources. It also sheds light onto the appearance of nonzero local spin in nonparaxial sound waves [32–34], which do not feature a polarization degree of freedom in the paraxial regime and correspond to spin-0 quantum particles (phonons).

**Acknowledgements:** We acknowledge help of Uwe Mick with the fabrication of samples. This work was partially supported by European Research Council (Starting Grant ERC-2016-STG-714151-PSINFONI and iCOMM Project No. 789340), EPSRC (UK), Excellence Initiative of Aix Marseille University — A*MIDEX, a French 'Investissements d'Avenir' programme, NTT Research, Army Research Office (ARO) (Grant No. W911NF-18-1-0358), Japan Science and Technology Agency (JST) (via the Q-LEAP program, and the CREST Grant No. JPMJCR1676), Japan Society for the Promotion of Science (JSPS) (JSPS-RFBR Grant No. 17-52-50023, and JSPS-FWO Grant No. VS.059.18N), the Foundational Questions Institute Fund (FQXi, Grant No. FQXi-IAF19-06), and a donor advised fund of the Silicon Valley Community Foundation.

**Author contributions:** K.Y.B. conceived the idea of this research, made theoretical calculations with input from M.A.A., and prepared the manuscript with input from all the authors.
*Focused-beam experiment:* P.B. and J.S.E. developed the idea of the experiment. J.S.E. performed the experiment. J.S.E. and P.B. performed the data processing. J.S.E. and P.B. wrote the corresponding part of the manuscript.
*Evanescent-wave experiment:* F.J.R.-F, D.J.R, L.H.N, and A.V.Z developed the idea of the experiment. D.J.R and L.H.N designed and performed the experiment. F.J.R.-F performed theoretical modeling. D.J.R and F.J.R.-F performed data processing. D.J.R. fabricated the samples. F.J.R.-F, D.J.R, L.H.N and A.V.Z wrote the related part of the manuscript.

# References

1. R. M. A. Azzam and N. M. Bashara, *Ellipsometry and Polarized Light* (North-Holland, Amsterdam, 1977).
2. V. B. Berestetskii, E. M. Lifshitz, and L. P. Pitaevskii, *Quantum Electrodynamics* (Pergamon, Oxford, 1982).
3. L. Novotny and B. Hecht, *Principles of Nano-Optics* (Cambridge University Press, Cambridge, 2012).
4. T. Carozzi, R. Karlsson, and J. Bergman, "Parameters characterizing electromagnetic wave polarization", *Phys. Rev. E* **61**, 2024–2028 (2000).
5. T. Setälä, A. Shevchenko, M. Kaivola, and A. T. Friberg, "Degree of polarization for optical near fields", *Phys. Rev. E* **66**, 016615 (2002).
6. M. R. Dennis, "Geometric interpretation of the three-dimensional coherence matrix for nonparaxial polarization", *J. Opt. A: Pure Appl. Opt.* **6**, S26–S31 (2004).
7. J. C. Petruccelli, N. J. Moore, and M. A. Alonso, "Two methods for modeling the propagation of the coherence and polarization properties of nonparaxial fields", *Opt. Commun.* **283**, 4457–4466 (2010).
8. C. J. R. Sheppard, "Jones and Stokes parameters for polarization in three dimensions", *Phys. Rev. A* **90**, 023809 (2014).
9. S. J. van Enk and G. Nienhuis, "Spin and orbital angular momentum of photons", *Europhys. Lett.* **25**, 497–501 (1994).
10. M. V. Berry and M. R. Dennis, "Polarization singularities in isotropic random vector waves", *Proc. R. Soc. A* **457**, 141–155 (2001).
11. K. Y. Bliokh, M. A. Alonso, E. A. Ostrovskaya, and A. Aiello, "Angular momenta and spin–orbit interaction of nonparaxial light in free space", *Phys. Rev. A* **82**, 063825 (2010).
12. R. P. Cameron, S. M. Barnett, and A. M. Yao, "Optical helicity, optical spin and related quantities in electromagnetic theory", *New J. Phys.* **14**, 053050 (2012).
13. K. Y. Bliokh and F. Nori, "Transverse and longitudinal angular momenta of light", *Phys. Rep.* **592**, 1–38 (2015).
14. K.Y. Bliokh and F. Nori, "Transverse spin of a surface polariton", *Phys. Rev. A* **85**, 061801(R) (2012).
15. P. Banzer, M. Neugebauer, A. Aiello, C. Marquardt, N. Lindlein, T. Bauer, G. Leuchs, "The photonic wheel – demonstration of a state of light with purely transverse angular momentum", *J. Eur. Opt. Soc. Rap. Publ.* **8**, 13032 (2013).
16. K. Y. Bliokh, A. Y. Bekshaev, and F. Nori, "Extraordinary momentum and spin in evanescent waves", *Nat. Commun.* **5**, 3300 (2014).
17. A. Canaguier-Durand and C. Genet, "Transverse spinning of a sphere in a plasmonic field", *Phys. Rev. A* **89**, 033841 (2014).
18. M. Neugebauer, T. Bauer, P. Banzer, and Gerd Leuchs, "Polarization Tailored Light Driven Directional Optical Nanobeacon", *Nano Lett.* **14**, 2546–2551 (2014).
19. F. J. Rodríguez-Fortuño, G. Marino, P. Ginzburg, D. O'Connor, A. Martinez, G. A. Wurtz, and A. V. Zayats, "Near-field interference for the unidirectional excitation of electromagnetic guided modes", *Science* **340**, 328–330 (2013).
20. J. Petersen, J. Volz, and A. Rauschenbeutel, "Chiral nano-photonic waveguide interface based on spin-orbit interaction of light", *Science* **346**, 67–71 (2014).
21. D. O'Connor, P. Ginzburg, F. J. Rodríguez-Fortuño, G. A. Wurtz, and A. V. Zayats, "Spin-orbit coupling in surface plasmon scattering by nanostructures", *Nat. Commun.* **5**, 5327 (2014).
22. M. Neugebauer, T. Bauer, A. Aiello, and P. Banzer, "Measuring the transverse spin density of light", *Phys. Rev. Lett.* **114**, 063901 (2015).
23. B. Le Feber, N. Rotenberg, and L. Kuipers, "Nanophotonic control of circular dipole emission", *Nat. Commun.* **6**, 6695 (2015).

24. Y. Lefier and T. Grosjean, "Unidirectional sub-diffraction waveguiding based on optical spin-orbit coupling in subwavelength plasmonic waveguides", *Opt. Lett.* **40**, 2890–2893 (2015).
25. K. Y. Bliokh, D. Smirnova, and F. Nori, "Quantum spin Hall effect of light", *Science* **348**, 1448 (2015).
26. A. Y. Bekshaev, K. Y. Bliokh, and F. Nori, "Transverse Spin and Momentum in Two-Wave Interference", *Phys. Rev. X* **5**, 011039 (2015).
27. T. Bauer, M. Neugebauer, G. Leuchs, and P. Banzer, "Optical Polarization Möbius Strips and Points of Purely Transverse Spin Density", *Phys. Rev. Lett.* **117**, 013601 (2016).
28. M. Neugebauer, J. S. Eismann, T. Bauer, and P. Banzer, "Magnetic and electric transverse spin density of spatially confined light" *Phys. Rev. X* **8**, 021042 (2018).
29. A. Aiello, P. Banzer, M. Neugebauer, and G. Leuchs, "From transverse angular momentum to photonic wheels", *Nat. Photon.* **9**, 789–795 (2015).
30. K. Y. Bliokh, F. J. Rodríguez-Fortuño, F. Nori, and A. V. Zayats, "Spin-orbit interactions of light", *Nat. Photon.* **9**, 796–808 (2015).
31. P. Lodahl, S. Mahmoodian, S. Stobbe, A. Rauschenbeutel, P. Schneeweiss, J. Volz, H. Pichler, and P. Zoller, "Chiral quantum optics", *Nature* **541**, 473–480 (2017).
32. C. Shi, R. Zhao, Y. Long, S. Yang, Y. Wang, H. Chen, J. Ren, and X. Zhang, "Observation of acoustic spin", *Natl. Sci. Rev.* **6**, 707–712 (2019).
33. K. Y. Bliokh and F. Nori, "Spin and orbital angular momenta of acoustic beams", *Phys. Rev. B* **99**, 174310 (2019).
34. I. D. Toftul, K. Y. Bliokh, M. I. Petrov, and F. Nori, "Acoustic Radiation Force and Torque on Small Particles as Measures of the Canonical Momentum and Spin Densities", *Phys. Rev. Lett.* **123**, 183901 (2019).
35. S. Golat, E. A. Lim, and F. J. Rodríguez-Fortuño, "Evanescent Gravitational Waves", arXiv:1903.09690 (2019).
36. K. Lindfors, T. Setälä, M. Kaivola, and A. T. Friberg, "Degree of polarization in tightly focused optical fields", *J. Opt. Soc. Am. A* **22**, 561–568 (2005).
37. K. Lindfors, A. Priimagi, T. Setälä, A. Shevchenko, A. T. Friberg, and M. Kaivola, "Local polarization of tightly focused unpolarized light", *Nat. Photon.* **1**, 228–231 (2007).
38. K. Y. Bliokh, A. Y. Bekshaev, and F. Nori, "Dual electromagnetism: helicity, spin, momentum and angular momentum", *New J. Phys.* **15**, 033026 (2013).
39. P. Banzer, U. Peschel, S. Quabis, and G. Leuchs, "On the experimental investigation of the electric and magnetic response of a single nano-structure", *Opt. Express* **18**, 10905–10923 (2010).
40. B. Richards and E. Wolf, "Electromagnetic diffraction in optical systems. II. Structure of the image field in an aplanatic system", *Proc. R. Soc. Lond. A Math. Phys. Sci.* **253**, 358–379 (1959).
41. J. S. Eismann, P. Banzer, and M. Neugebauer, "Spin-orbit coupling affecting the evolution of transverse spin", *Phys. Rev. Research* **1**, 033143 (2019).